\documentstyle[useAMS,times,psfig,graphics,astrobib,amssymb]{mn2e}

\newcommand{\be}{\begin{equation}}
\newcommand{\e}{\end{equation}}
\newcommand{\bear}{\begin{eqnarray}}
\newcommand{\ear}{\end{eqnarray}}

\begin{document}

\title[A very extended reionization epoch ?]
{A very extended reionization epoch ?}
\author[Melchiorri, Choudhury,  Serra, Ferrara]
{A. Melchiorri$^1$\thanks{E-mail: alessandro.melchiorri@roma1.infn.it},
T. Roy Choudhury$^2$\thanks{E-mail: chou@sissa.it},
P. Serra$^1$\thanks{E-mail: paolo.serra@roma1.infn.it},
A. Ferrara$^2$\thanks{E-mail: ferrara@sissa.it}\\
$^1$Dipartimento di Fisica e Sezione INFN, Universita' di Roma
``La Sapienza'', Ple Aldo Moro 5, 00185, Roma, Italy\\
$^2$SISSA/ISAS, via Beirut 2-4, 34014 Trieste, Italy
}

\maketitle

\date{\today}

\begin{abstract}

The recent observations of cross temperature-polarization
power spectra of the Cosmic Microwave Background (CMB) made by
the WMAP satellite are in better agreement with a high value
of the Thomson scattering optical depth $\tau \approx 0.17$.
This value is close to $\tau = 0.3$, which is taken as the upper limit 
in the parameter extraction analysis made by the WMAP team.
However, models with $\tau \sim 0.3$ 
provide a good fit to current CMB data and are not significantly 
excluded when combined with Large Scale Structure data. 
By making use 
of a self-consistent reionization model, we verify the 
astrophysical feasibility of models with $\tau \sim 0.3$. 
It turns out that current data 
on various observations related to the thermal and 
ionization history of the intergalactic medium are not 
able to rule out $\tau \sim 0.3$.
The possibility of a very extended reionization epoch can 
significantly undermine the WMAP constraints on crucial cosmological 
parameters such as the Hubble constant, the spectral index of
primordial fluctuations and the amplitude of dark matter clustering. 
\end{abstract}

\section{Introduction}

The recent results on Cosmic Microwave Background (CMB) anisotropy 
from the WMAP satellite \cite{wmap1,spergel}
represent an extraordinary success for the 
standard cosmological model of structure formation based on 
Cold Dark Matter (CDM) and adiabatic primordial perturbations.  
Furthermore, data releases from the Sloan Digital Sky Survey 
(SDSS; \citeNP{sloan}) are living up to expectations and combined analysis 
of all these datasets are placing strong constraints on 
most cosmological parameters. 
 
However, even if theory and observations are in spectacular  
agreement, discrepancies seem to be present and have already stimulated
the interest of several authors. One of most unexpected results
from the first year WMAP data is indeed the detection of an excess in the
large angular scale cross temperature-polarization power
spectrum \cite{kogut}. In the standard scenario, this excess may be interpreted
as due to a late reionization process which produces a 
second last scattering surface at lower redshifts ($\sim
10-20$).  The CMB radiation can therefore get polarized through 
anisotropic Thomson scattering on angular scales comparable to the
horizon at reionization ($\theta > 5^{\circ}$).

Analysis made by the WMAP team of the cross temperature-polarization
data constrains the best-fit value of the Thomson optical depth of the universe
to $\tau \approx 0.17$ (the exact best-fit value and the errors depend
on the analysis technique employed).
Following this, considerable effort is spent in
constructing models, both numerical \cite{cfw03,ro04,sahs03,syahs04,gnedin04}
and semi-analytical \cite{wl03a,hh03,cf05}, analyzing the  
feasibility of such a high $\tau$.  
In this context, it was shown that an early population 
of metal-free stars, with reasonably low values of star-forming
efficiency and escape fraction \cite{cfw03,cf05}, 
is sufficient to produce such a high $\tau$.

In this paper we want to emphasize another aspect of this result and
investigate the compatibility of cosmological and astrophysical data
with a very extended reionization process i.e. $\tau~\sim~0.3$.
As we illustrate in the next section, the parameter extraction made
by the WMAP team has not considered this case, since a top-hat prior 
of $\tau < 0.3$ has been included in their analyses (see
\citeNP{spergel} and \citeNP{peiris}). Given the fact
that the reionization process is still to be fully understood, 
one could wonder about the robustness of this assumption.
In the next section we will indeed show that $\tau=0.3$ is 
compatible with the data and, as already pointed out in 
\cite{tegmark2,fosalba}, letting $\tau$ to vary freely
up to $\tau =0.5$ more than doubles the error bars on cosmological parameters.
Studying the astrophysical feasibility of a very extended
reionization is therefore not only important for the 
theoretical understanding of the process but also for
determining the cosmological parameters with highest
possible accuracy and reliability.

The paper is organized as follows: In the next section, we 
analyze current CMB and Large Scale Structure data to show that 
high values of the optical depth are not 
excluded and discuss the implications of having $\tau \sim 0.3$ 
on cosmological parameters. 
In section 3, we verify whether models with 
$\tau \sim 0.3$ violate any observational constraints related 
to the ionization and thermal history of the universe. 
The final section presents our conclusions.

\section{Cosmological Constraints on the optical depth}

\begin{figure*}
\rotatebox{270}{\resizebox{0.42\textwidth}{!}{\includegraphics{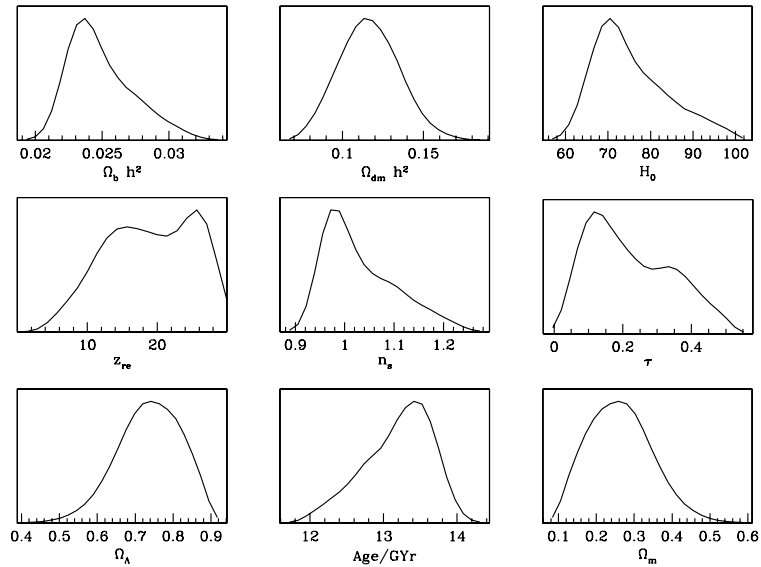}}}
\caption{Likelihood distribution functions on the cosmological
  parameters described in the text. Only the WMAP data is used
without prior on the optical depth. As we can see, values of
$\tau \sim 0.3$ are well consistent with the data}
\label{wmapsolo}
\end{figure*}

\begin{figure}
\rotatebox{270}{\resizebox{0.32\textwidth}{!}{\includegraphics{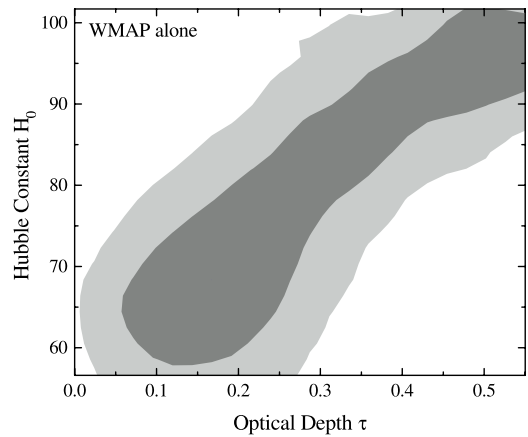}}}
\caption{$1$-$\sigma$ and $2$-$\sigma$ likelihood contour plot in the $H_0$ vs $\tau$
plane from the WMAP data alone. There is a clear degeneracy
between the two parameters and the Hubble parameter is
unconstrained from the CMB data alone. Including the prior $\tau <0.3$
artificially constrains the Hubble parameter.}
\label{Hubble}
\end{figure}

In this Section we investigate the possibility of having a high
optical depth $\tau \sim 0.3$ by analyzing current
CMB and Large Scale Structure data.
The method we adopt is based on the Markov Chain Monte Carlo package
\texttt{cosmomc}\footnote{Available from
\texttt{http://cosmologist.info}.} \cite{Lewis:2002ah}.
We use individual chains and we adopt the Raftery and Lewis convergence
diagnostics\footnote{see \texttt{http://cosmologist.info
/cosmomc/readme.html}.}
We sample the following 6 dimensional set of cosmological parameters,
adopting flat priors on them: the physical baryon
and CDM densities, $\omega_b \equiv \Omega_b h^2$ and
$\omega_{\rm dm} \equiv \Omega_{\rm dm} h^2$, 
the ratio of the sound horizon to the
angular diameter distance at decoupling, $\theta_s$, the scalar
spectral index and the overall normalization of the spectrum,
$n_s$ and $A_s$, and, finally, the optical depth to reionization,
$\tau$. Furthermore, we consider purely adiabatic initial
conditions, we impose flatness and we do not include gravitational
waves. Relaxing flatness does not affect 
$\tau$ in a very significant way; however, it does alter the values of the 
other basic parameters and thus, may further change the
interpretation and astrophysical implications of the overall results.
We restrict our analysis to the case of 3 massless neutrino
families.
We include the first-year temperature and
polarization data \cite{wmap1} with the routine for computing the likelihood
supplied by the WMAP team \cite{verde}  as well as the CBI
\cite{CBI} and ACBAR measurements of the CMB.

In addition to the CMB data, we also consider the constraints on the
real-space power spectrum of galaxies from the SDSS \cite{sloan}. 
We restrict the analysis to a
range of scales over which the fluctuations are assumed to be in
the linear regime ($k < 0.2 h^{-1}\rm Mpc$). When combining the
matter power spectrum with CMB data, we marginalize over a bias
$b$, constant with scale and redshift, considered as an 
additional nuisance parameter. 

The results of our analysis are reported in Table 1 and 
displayed in Figures 1--4.
In Figure 1 we show the likelihood distribution function obtained
by analyzing the WMAP data only. As we can see the data does not 
provide strong constraints on $\tau$ and values as large as 
$\tau \sim 0.5$ are compatible with the data. 
As a further example, we plot in Figure 2 the $2-D$ likelihood distribution
on the $H_0$-$\tau$ plane after marginalization over the remaining
{\it nuisance} parameters. As we can see there is a clear degeneracy
between the two parameters and the value of the Hubble constant is
not constrained by the WMAP data alone. The claimed constraint in
\citeN{spergel} of $h=0.72 \pm 0.05$ at $68 \%$ c.l. is mostly due to
the top prior on $\tau$.
For comparison, we report in Table 1 the results from 
a similar analysis but with now the inclusion of an external top-hat prior 
of $\tau < 0.3$ as in the analysis made by the WMAP team \cite{spergel}.  
As we can see, the inclusion of the optical depth prior 
more than halves the error bars and greatly improves the constraints 
on the Hubble parameter, the spectral index and the baryon density to 
values in agreement with those reported by the WMAP team.

It is useful to compare the WMAP constraints 
derived in the case of  $\tau\sim0.30$ with independent 
cosmological probes.
If we assume a high value of $\tau=0.30\pm0.01$, then 
the likelihood best fit 
values are $\omega_b =0.026$, $h= 0.78$ and $n_s= 1.07$.
A value of $h\sim0.8$ is still in reasonable agreement with the 
HST constraint of $h=0.72\pm0.07$ \cite{freedman}. Values of 
$\omega_b > 0.025$ are difficult to reconcile with abundances of
primordial light elements and standard Big Bang Nucleosynthesis
which provide $\omega_b=0.020\pm0.001$ \cite{bbn}.
However, neutrino physics is still rather unknown and
deviations from the standard model, like inclusion of
 neutrino chemical potential or extra relativistic degrees
of freedom could be present and larger values of 
$\omega_b$ may be possible (see e.g. \citeNP{hansen}).
We can therefore conclude that the high $\tau$ case 
is in agreement with the WMAP data and that the 
best fit cosmological model with $\tau=0.3$ is 
in reasonable agreement with several other observables.

\begin{figure*}
\rotatebox{270}{\resizebox{0.42\textwidth}{!}{\includegraphics{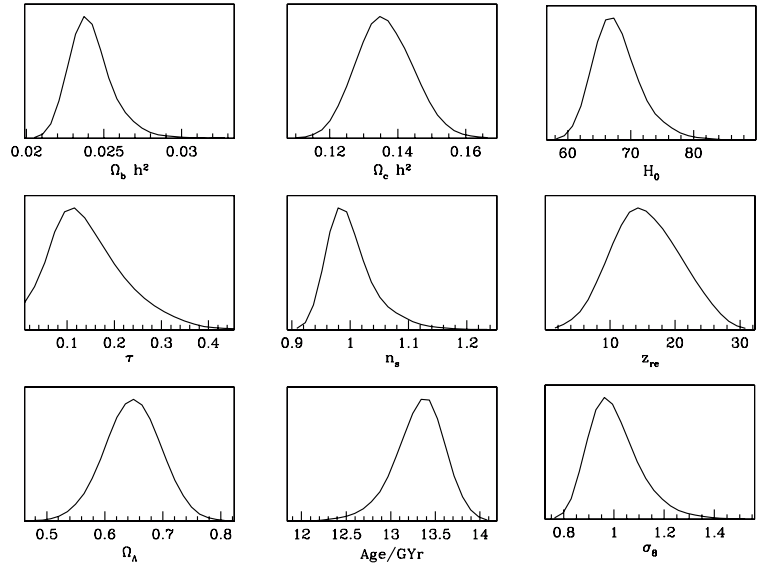}}}
\caption{Likelihood distribution functions on the cosmological
  parameters described in the text. The WMAP+SDSS data is used
without external prior on the optical depth. Values of $\tau \sim 0.3$
are now less in agreement with the data but still inside the $2
  \sigma$ level.}
\label{wmapsloan}
\end{figure*}

\begin{figure*}
{\resizebox{0.28\textwidth}{!}{\includegraphics{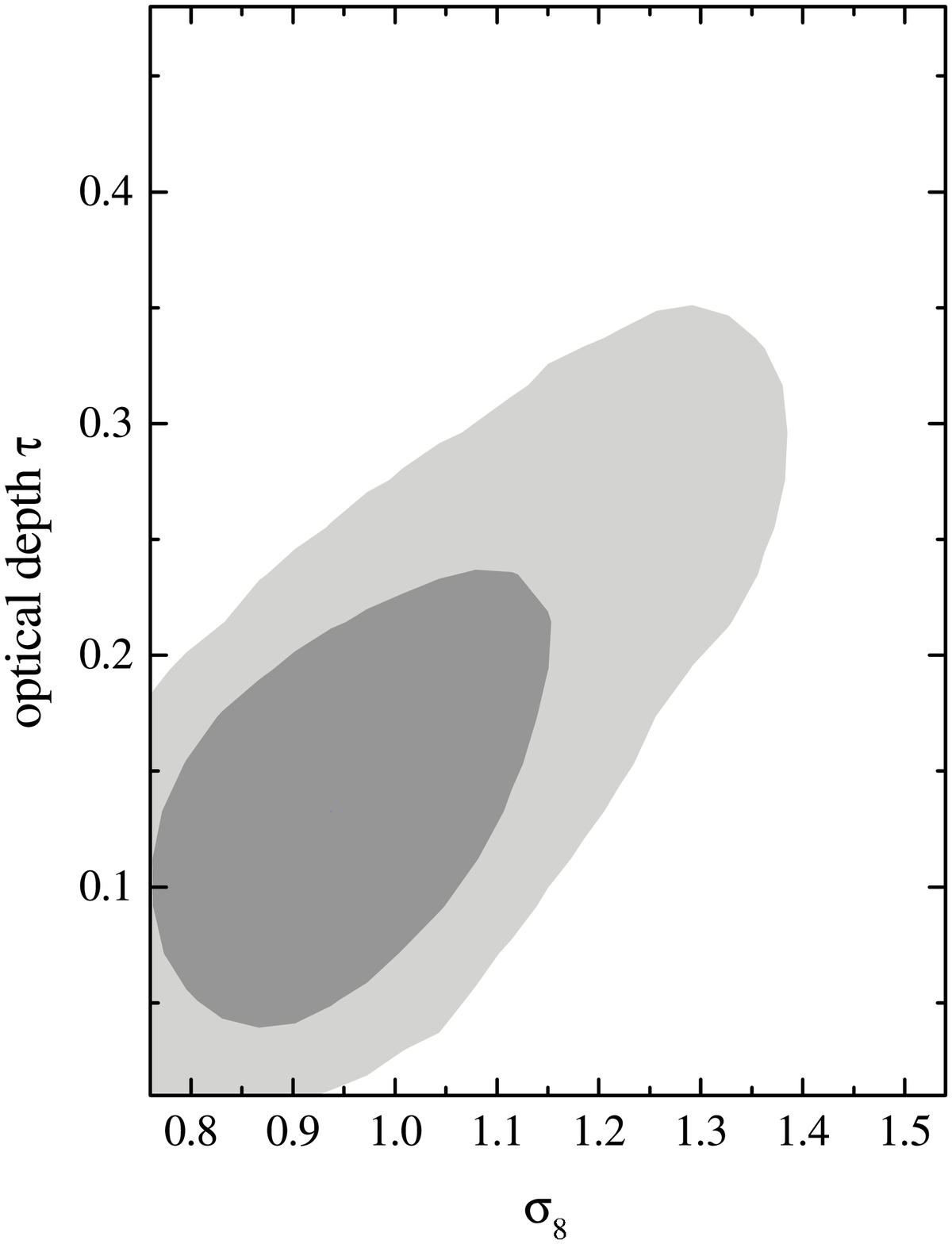}}}
~~~~~~~~~~~~~
{\resizebox{0.28\textwidth}{!}{\includegraphics{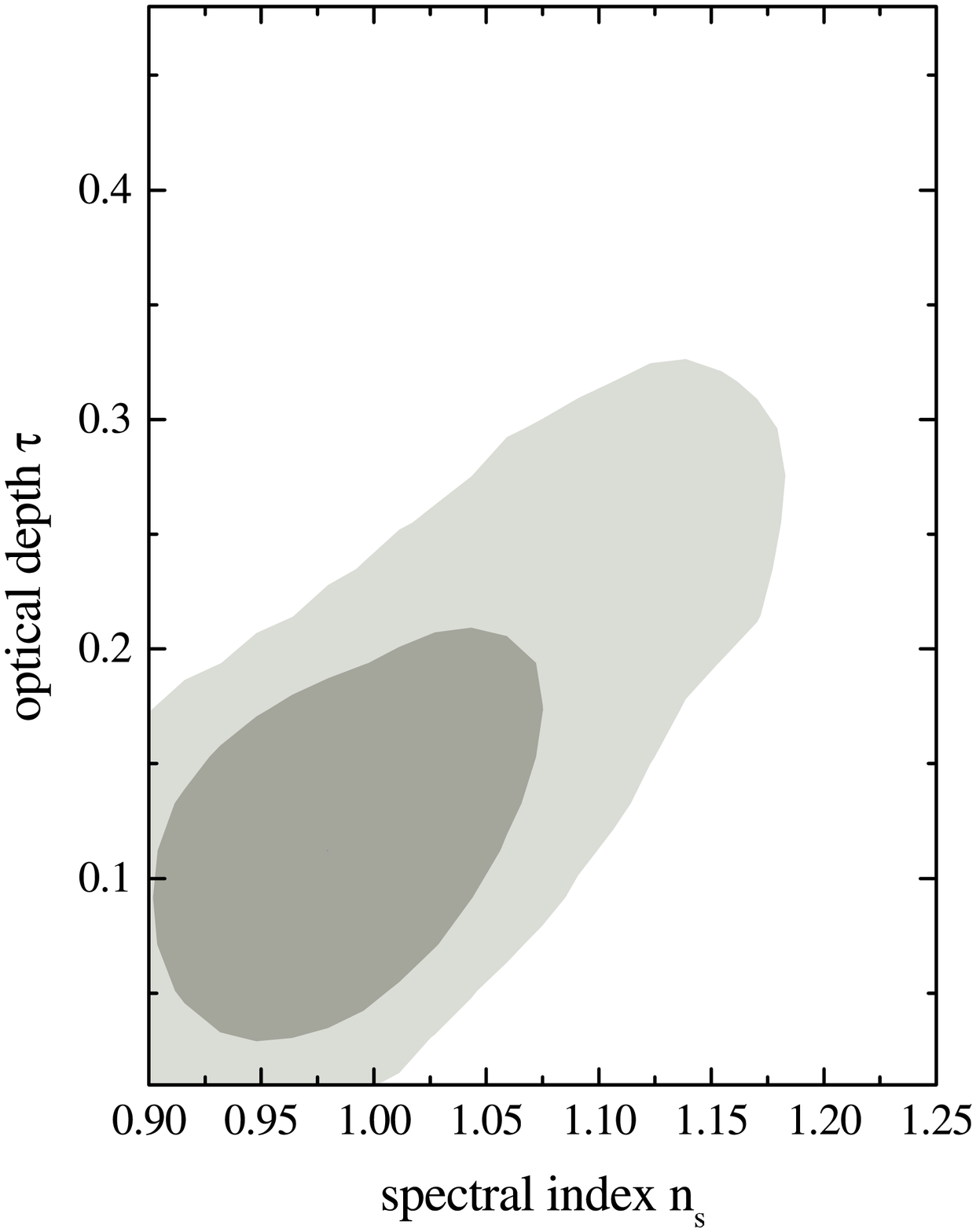}}}
\caption{$1$-$\sigma$ and $2$-$\sigma$ likelihood contour plots in the
  $\tau$ vs $\sigma_8$ (left panel) and $\tau$ vs $n_s$ (right panel)
planes from a WMAP+SDSS analysis. Even after the inclusion of
large scale structure data, there is a clear degeneracy
between the parameters. Values of $\tau \sim 0.3$ are in between the
$2$ $\sigma$ contours if $n_s\sim1.1$ and $\sigma_8\sim 1.17$.}
\label{sigmaotto}
\end{figure*}

Including the SDSS data, as we do in Figure 3, breaks
several degeneracies, mainly between the spectral index $n_s$
and $\tau$, and improves the constraints. However, as we can see from
Figure 4, models with $\tau=0.3$ are still inside the 
2-$\sigma$ level and are not strongly disfavoured. 
In the WMAP+SDSS analysis,
the best fit model under the assumption of $\tau=0.30\pm0.01$ has 
$\omega_b = 0.027$, $h= 0.75$, $n_s= 1.08$ and $\sigma_8=1.17$.
These results are close to those from the previous WMAP 
analysis further confirming the reasonable agreement 
between the SDSS data and $\tau=0.3$.

Including ACBAR and CBI experiments to WMAP+SDSS
improves the constraints on $\tau$ and we get an upper limit
of $\tau \le 0.25$. While this disfavours 
the $\tau \sim 0.3$ hypothesis, one should also note
that other effects like massive neutrinos (see \citeNP{fogli}), 
running spectral 
index and a scale-dependent bias, just to name a few, 
could all be present and improve the agreement 
between models with high $\tau$ and the 
data. For example, including
a running spectral index in the primordial spectrum, which may be
motivated by inflation and/or by a scale
dependent bias in galaxy clustering, enlarges 
the constraints again to $\tau \le 0.32$.
In the case of running spectral index and $\tau =0.30$ the best fit 
parameters are $\omega_b = 0.024$, $h= 0.67$, $n_s= 0.93$,
${\rm d}n/{\rm d} \ln k=-0.08$ and $\sigma_8=1.11$.

An accurate determination  of $\sigma_8$ may be the
best way to determine $\tau$.
Unfortunately there is a large scattering in the constraints on 
$\sigma_8$ from current observations.
A value of $\sigma_8 =  1.1-1.2$ is compatible with
the determinations inferred from the local cluster X-ray 
temperature function (see e.g. \citeNP{pierpa,borgani,eke}) 
and cosmic shear data \cite{vande,refregier,bacon}, but 
incompatible with other analysis (see e.g. \citeNP{s8eljak,liddle}).
Recent measurements of galaxy clustering at redshift $z=0.6$ 
from the COMBO-17 survey gives $\sigma_8=1.02\pm0.17$ at zero
redshift \cite{peacock}. It is interesting to point out 
that, if the small-scale CMB anisotropy excess measured by the CBI experiment 
(see \citeNP{CBI}) is indeed 
due to integrated Sunyaev-Zeldovich effect, this would be
compatible with a value of $\sigma_8=1.04\pm0.12$ \cite{komatsu}
therefore in agreement with a very extended reionization scenario.

\begin{table*}
\caption
{Mean values and marginalized $95 \%$ c.l. limits for several cosmological 
parameters from WMAP and SDSS (see text for details).}
\begin{center}
\begin{tabular}{|c|c|c|c|c|c|}
\hline
Parameter & WMAP only & WMAP + $\tau<0.30$ & WMAP + $\tau=0.30\pm0.01$ & WMAP + SDSS & WMAP + SDSS + $\tau=0.30\pm0.01$ \\
\hline
$\Omega_bh^2$&$0.025_{-0.004}^{+0.006}$&$0.024_{-0.003}^{+0.002}$&$0.026_{-0.002}^{+0.002}$&$0.024_{-0.002}^{+0.004}$&$0.026_{-0.001}^{+0.002}$ \\
$\Omega_{\rm dm}h^2$&$0.12_{-0.04}^{+0.03}$&$0.12_{-0.03}^{+0.03}$&$0.11_{-0.03}^{+0.03}$&$0.14_{-0.02}^{+0.01}$&$0.13_{-0.01}^{+0.02}$\\
$h$&$0.75_{-0.12}^{+0.21}$&$0.71_{-0.08}^{+0.09}$&$0.79_{-0.08}^{+0.04}$&$0.68_{-0.06}^{+0.08}$&$0.74_{-0.05}^{+0.05}$\\
Age&$13.2_{-1.0}^{+0.7}$&$13.4_{-0.5}^{+0.5}$&$13.0_{-0.4}^{+0.4}$&$13.3_{-0.6}^{+0.5}$&$12.9_{-0.6}^{+0.4}$\\
$n_s$&$1.03_{-0.10}^{+0.16}$&$0.99_{-0.06}^{+0.07}$&$1.06_{-0.04}^{+0.05}$&$1.00_{-0.06}^{+0.11}$&$1.08_{-0.04}^{+0.12}$\\
$\tau$&$0.22_{-0.18}^{+0.25}$&$0.15_{-0.10}^{+0.12}$&$0.30_{-0.02}^{+0.02}$&$0.15_{-0.10}^{+0.15}$&$0.29_{-0.01}^{+0.03}$\\
$\sigma_8$& -- & -- & -- &$1.00_{-0.15}^{+0.26}$&$1.18_{-0.11}^{+0.14}$\\

\hline
\end{tabular}
\label{table:modpar}
\end{center}
\end{table*}

\section{Astrophysical Constraints}

A potential challenge to models with high values 
of $\tau$ can come from the astrophysical constraints. To address
this issue, we study the implications of a high value 
of $\tau \sim 0.3$ 
on reionization history of the universe. For this purpose, we use 
a semi-analytical model \cite{cf05} developed for studying cosmic
reionization and thermal history of the intergalactic medium (IGM). 
The model implements most of
the relevant physics governing these processes, such as the
inhomogeneous IGM density distribution, three different classes of
ionizing photon sources (massive PopIII stars, PopII stars and QSOs),
and radiative feedback inhibiting star formation in low-mass
galaxies. The main advantage of the model is that it can be constrained
quite well by comparing its predictions 
with various observational data, namely, 
the redshift evolution of Lyman-limit absorption systems \cite{smih94},
Gunn-Peterson \cite{songaila04} and electron scattering optical depths
\cite{kogut} and cosmic star formation history \cite{nchos04}; 
in addition the model also includes constraints obtained by
\citeN{sf03} on Near Infra Red
Background (NIRB) data \cite{mcf++00,gwc00,crbj01,wright01}.
In this paper, we extend the above model of \citeN{cf05} to include
the possibility of a high $\tau$ along with high values 
of $n_s = 1.08, \sigma_8 = 1.18$ and $\omega_b = 0.026$, 
as discussed in the previous section and reported in Table 1.

The most crucial difficulties in admitting 
such a high value of $\tau$ within
standard reionization scenarios can be divided
into two parts.
In the first part, we discuss whether it is possible to have sources 
which can produce a high number 
of ionizing photons so as to produce a large $\tau$. 
A common mechanism for obtaining a large $\tau$ is through 
metal-free PopIII stars, which has been 
used widely (see, for example, \citeNP{cen03a,cen03b,vts03}), and is 
in agreement with various set of observations \cite{cf05}
including the NIRB data \cite{sf03,msf03}.
In such scenarios,  PopIII stars with a 
photon production efficiency $\epsilon_{\rm III} 
\equiv \epsilon_{\rm SF, III} f_{\rm esc, III} \approx 0.006$ 
(where $\epsilon_{\rm SF, III}$ is the star-forming efficiency
and $f_{\rm esc, III}$ is the escape fraction of ionizing photons) 
are able to 
produce a $\tau = 0.17$ (the best-fit value obtained by 
fitting the temperature-polarization 
cross power spectrum to $\Lambda$CDM models in 
which all parameters except $\tau$ assume their best fit values based on 
the temperature power spectrum; \citeNP{kogut}).
One might then expect that a 
considerable higher value 
of $\epsilon_{\rm III}$ must be required to produce a $\tau$ of 0.3. 
However, one should also keep in mind that 
the model with $\tau = 0.3$ has higher 
values of $\sigma_8 = 1.18$ and $\omega_b = 0.026$ 
compared to the standard one, which means that 
(i) the number of collapsed haloes is higher and 
(ii) the baryonic mass within the collapsed haloes is larger. 
In fact, the value of $\epsilon_{\rm III}$ required to match 
the high $\tau$ is $\approx 0.0035$, which 
is similar to the value in standard scenarios. 
So we 
conclude that production 
of a large number of ionizing photons at high redshift does not 
pose any serious difficulty for models with high $\tau$.

Various other mechanisms for early reionization have been suggested 
in order to produce a high $\tau$, most notably being 
an early population of accreting black holes
\cite{mrvho04,dhl04,ro04b}. The accretion 
of gas onto these black holes or miniquasars would 
produce a background of hard photons, which 
have a large mean free path and can ionize large regions 
of the IGM. However, the constraints from the present-day soft X-ray
background imply that these sources can, at most, produce 
$\sim 3$ photons per hydrogen atom \cite{dhl04,shf05}, 
which is much less than the 
requirement for fully ionizing the IGM -- hence 
the reionization from these sourced would be partial. 
While the nature of reionization (topology of the 
ionized regions, thermal state of the IGM etc) would 
be quite different if it is achieved by the black holes rather than 
PopIII stars, 
whether the photons from the miniquasars can produce 
a high value of $\tau \approx 0.3$ without violating any other 
observational constraints needs to be verified.

\begin{figure}
\rotatebox{270}{\resizebox{0.4\textwidth}{!}{\includegraphics{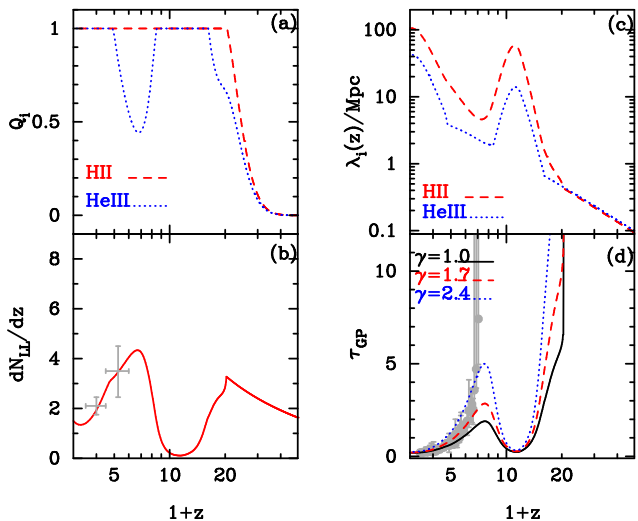}}}
\caption{Comparison of a model having $\tau = 0.3$ with various 
astrophysical observations. 
Adopted parameters are $\epsilon_{\rm III}= 3.5\times 10^{-3}$, $\epsilon_{\rm II}=3\times 10^{-4}$,
$z_{\rm trans}=11$.  The panels show as a function of redshift: 
(a) filling factor of ionized hydrogen and doubly-ionized helium regions, 
(b) specific number of Lyman-limit systems, 
(c) ionizing photons mean free path for hydrogen and helium,  
(d) Lyman-$\alpha$ Gunn-Peterson optical depth for three
values of slope $\gamma$ of the IGM temperature-density relation.
}
\label{zt11}
\end{figure}

The second difficulty regarding high values of $\tau$ is whether
such a high number of photons at high redshift violate any observational
constraints (see Figure \ref{zt11}).
The first set of such constraints come from the observations related
to the IGM.
Since there are virtually no
observational data on IGM at $z > 6$, the most severe constraints
on the model comes from the Lyman-$\alpha$ Gunn-Peterson optical depth 
$\tau_{\rm GP}$ at $z \approx 6$. In the standard scenario, 
$\tau_{\rm GP}$ observations are very well matched by the flux from 
normal PopII stars with an efficiency 
$\epsilon_{\rm II} \approx 0.0005$ 
along with a population of QSOs as required 
by data on optical luminosity function. 
It is thus necessary that the PopIII stars do not have a significant 
flux at $z \approx 6$. 
For the models studied in this paper, 
this provides a mild constraint on the transition 
redshift of PopIII stars $z_{\rm trans} \gtrsim 10.5$.
However, there is 
another physical process which affects $\tau_{\rm GP}$ 
and thus provides further constraints.
This has to do with the mean free path $\lambda_H$ of 
hydrogen-ionizing photons.
At high redshifts, the high flux from PopIII stars create huge ionized 
regions, thus producing high values of $\lambda_H \sim$ 60--80 Mpc. 
Once the PopIII stars disappear at $z_{\rm trans}$, these ionized
regions start becoming neutral over a 
recombination time-scale resulting in a 
gradual decrease of $\lambda_H$. However,
the constraints on $\tau_{\rm GP}$ (taking into
account the uncertainty in the value of the 
slope of the IGM temperature-density relation $\gamma$) imply that 
$\lambda_H \lesssim$ few Mpcs at $z \approx 6$. 
To obtain such low values of $\lambda_H$, one has to make sure that 
the PopIII stars start disappearing around $z_{\rm trans} \approx 11$.
Once this condition is satisfied, the model is consistent 
with various available observations, namely, the redshift 
evolution of Lyman-limit systems, the 
temperature corresponding to the mean IGM density, the 
cosmic SFR and $\tau_{\rm GP}$,
as shown in Figure \ref{zt11}.

The second set of observational consequence is related to 
the NIRB which is believed to be due to the PopIII stars. 
The observations of NIRB constrain the combination
$\epsilon_{\rm SF, III} (1 - f_{\rm esc, III})$ and 
$z_{\rm trans}$ \cite{sf03,msf03}. 
First, note that the value of 
$\tau$ is sensitive only to the combination 
$\epsilon_{\rm SF, III} f_{\rm esc, III}$; hence, it is not difficult to 
match the NIRB constraints on 
$\epsilon_{\rm SF, III} (1 - f_{\rm esc, III})$
by suitably choosing the values 
of $\epsilon_{\rm SF, III}$ and $f_{\rm esc, III}$. Second,
the decline of the NIRB at wavelengths below the J band
constrains $z_{\rm trans} \approx 10$, whereas the $\tau_{\rm GP}$ constraints
at $z \gtrsim 6$ require that $z_{\rm trans} > 11$ (as discussed in the 
previous paragraph).
However, the constraint from NIRB is not too strong mainly 
due to lack of reliable data at wavelengths shortwards of  the J band.
It is thus difficult to rule out
models where the PopIII star formation rate
starts decreasing gradually from $z \approx 11$ with substantial
contribution remaining till $z \approx 9$.

In conclusion, current observations, related to the IGM and NIRB, 
are not able to rule out models with $\tau \sim 0.3$, though
they are only marginally consistent in some cases.

\section{Conclusions}

In this paper we have investigated the feasibility of a very
extended reionization. We can summarize our results in

\begin{itemize}

\item The first year WMAP data is in perfect agreement with an Thomson 
scattering optical depth parameter as high as $\tau =0.5$. Not considering this possibility
  in parameter extraction analysis may bias the results
on several quantities like the Hubble constant, the baryon
density and $\sigma_8$.

\item Including galaxy clustering data from SDSS disfavours models 
with $\tau =0.3$ but not at high significance. 
Similarly, inclusion of the CMB data from CBI and ACBAR improves 
the constraints to $\tau \le 0.25$. 
On the other hand, if additional
parameters like a running spectral index 
or scale dependent bias are considered,
high $\tau$ models are found to be in better agreement with observations.

\item We have therefore  studied the astrophysical feasibility of such a 
high value of $\tau$, particularly with respect to the 
observations related to the reionization history of the universe.
Using a simple semi-analytical model \cite{cf05}, we find that 
an early population of massive metal free PopIII stars with 
photon production efficiency $\epsilon_{\rm III}
\equiv \epsilon_{\rm SF,III} f_{\rm esc,III} 
\approx 0.0035$ is sufficient
to produce such a high $\tau$. However, in order that 
the photon flux does not violate the Lyman-$\alpha$ 
Gunn-Peterson optical depth
constraints at $z \gtrsim 6$, the PopIII star formation rate 
should start decreasing around $z_{\rm trans} \approx 11$. 
This value of $z_{\rm trans}$ 
is marginally consistent with the observations of NIRB.

\end{itemize}

We can therefore conclude that the possibility of a very extended
reionization epoch is still not completely ruled out by current
data and should be taken into account, especially when deriving reliable
 $2 \sigma$ constraints from CMB anisotropy measurements. 
In future, more accurate CMB polarization measurements, such as those 
expected 
from the Planck satellite experiment, together with high accuracy 
measurements of $\sigma_8$ from weak lensing and galaxy surveys, 
will be able to verify or rule out this interesting possibility.

\section*{Acknowledgments}

The authors would like to thank thank C. Baccigalupi and S. Leach for useful
discussions. AM work is part of the GEMINI-SZ project funded by MURST through
 COFIN contract no. 2004027755.

\end{document}